\documentclass[aps,prmaterials,reprint,amsmath,amssymb,floatfix,superscriptaddress]{revtex4-2}

\usepackage[utf8]{inputenc}
\usepackage[T1]{fontenc}
\usepackage{textcomp}
\usepackage{newunicodechar}
\newunicodechar{−}{\ensuremath{-}}
\usepackage{graphicx}
\usepackage[usenames,dvipsnames]{xcolor}
\usepackage[normalem]{ulem}
\usepackage{hyperref}
\hypersetup{colorlinks=true, urlcolor=blue, citecolor=blue, linkcolor=blue}

\definecolor{alicolor}{RGB}{0,115,115}

\begin{document}

\title{First-principles phase diagram of monolayer CrS$_2$ polymorphs under biaxial
strain and its sensitivity to the Hubbard U parameter}

\author{Rashmi Ranjan Routaray}
\email{rrr.phy@gmail.com}
\affiliation{Physique Th\'{e}orique des Mat\'{e}riaux, Q-MAT, Universit\'{e} de Li\`{e}ge,
B-4000 Sart-Tilman, Belgium}

\author{Ali Ghojavand}
\affiliation{Physique Th\'{e}orique des Mat\'{e}riaux, Q-MAT, Universit\'{e} de Li\`{e}ge,
B-4000 Sart-Tilman, Belgium}

\author{Eric Bousquet}
\email{eric.bousquet@uliege.be}
\affiliation{Physique Th\'{e}orique des Mat\'{e}riaux, Q-MAT, Universit\'{e} de Li\`{e}ge,
B-4000 Sart-Tilman, Belgium}

\date{\today}

\begin{abstract}
In correlated two-dimensional magnets the balance between competing electronic and magnetic phases is often set by the strength of the on-site Coulomb interaction, so that the predicted ground state can hinge on a parameter that is not uniquely defined. Monolayer CrS$_2$ lies squarely in this regime, hosting several energetically close polymorphs whose character ranges from non-magnetic semiconducting to magnetically ordered metallic. We therefore investigate, using DFT$+U$ within the generalized gradient approximation, the
structural, magnetic, and electronic properties of the 1H, 1T, 1T$'$, and 1T$''$ polymorphs of monolayer CrS$_2$ under varying Hubbard interactions ($U=0$--$8$~eV) and biaxial strain ($-10\%$ to $+10\%$).
The 1T-derived structures remain metallic and magnetic throughout, with $U$ acting mainly as a spectral renormalization. The 1H phase behaves differently, passing from a non-magnetic semiconductor through an
antiferromagnetic semiconducting regime before metallizing, so that the gap closes as $U$
increases while the local Cr moment grows monotonically. Biaxial tensile strain at fixed $U$ induces a similar sequence of transitions. Along both tuning pathways, the two inequivalent nearest-neighbor Cr--Cr exchange channels progressively separate and cross, with the parallel-spin channel becoming dominant, accompanied by a breathing-type bond disproportionation. Independent calculations with a localized-orbital implementation reproduce the overall evolution but yield shifted critical values of $U$ and of strain, demonstrating the robustness of the phase diagram while highlighting the implementation dependence of the transition thresholds.
\end{abstract}

\maketitle

\section{Introduction}
\label{sec:intro}
The ability to control, tune, and spatially pattern electronic and magnetic phases in atomically thin materials is a key objective for next-generation nanoelectronic technologies.
Two-dimensional (2D) transition-metal dichalcogenides (TMDs) provide a versatile platform for achieving this goal, offering opportunities for band-structure engineering, spin manipulation, and strain-controlled functionalities within a single atomic layer~\cite{Wang2012,Manzeli2017}. Beyond conventional single-particle band engineering~\cite{Basov2017}, electron--electron interactions introduce additional degrees of freedom, enabling metal--insulator transitions~\cite{Imada1998}, competing magnetic orders~\cite{Huang2021}, and strong magnetoelastic coupling between electronic, spin, and lattice degrees of freedom~\cite{Jang2023,Georges1996}. The interplay between these effects provides opportunities for controlling phase transitions and stabilizing distinct electronic and magnetic states, with potential implications for nanoscale spintronic and electronic devices~\cite{Guo2024QuantumSpintronics,Xu2024MagneticTransition,Ryee2026TMDMottMetal}.

Monolayer chromium disulfide (CrS$_2$) provides an interesting platform for exploring these effects. As a Cr-based dichalcogenide with partially filled 3d states and strong Cr--S hybridization, it exhibits several low-energy polymorphs with markedly different electronic and magnetic properties, ranging from non-magnetic (NM) semiconducting to magnetically ordered metallic states~\cite{Habib2019,Chen2021}. Among the proposed structures are the trigonal-prismatic 1H (also referred to as 2H) and octahedral 1T phases, as well as the 1T$'$ and 1T$''$ derivatives associated with dimerization and trimerization of the Cr sublattice, respectively. An additional orthorhombic polymorph has recently been predicted~\cite{Lei2024}, while the coexistence of the 1H, 1T, and 1T$'$ phases has been experimentally observed in chemical-vapour-deposited monolayer CrS$_2$~\cite{Habib2019}. More recently, catalyst-free growth of layered 1T-CrS$_2$ has been reported, revealing ferromagnetism above room temperature, an out-of-plane magnetic easy axis, and a topological Hall response at low temperatures~\cite{Obaidulla2026}. Given the energetic proximity of these competing phases, relatively small external perturbations may be sufficient to drive substantial changes in the structural, electronic, and magnetic ground states of CrS$_2$.

\begin{figure*}[!t]
  \centering
  \includegraphics[width=1\textwidth]{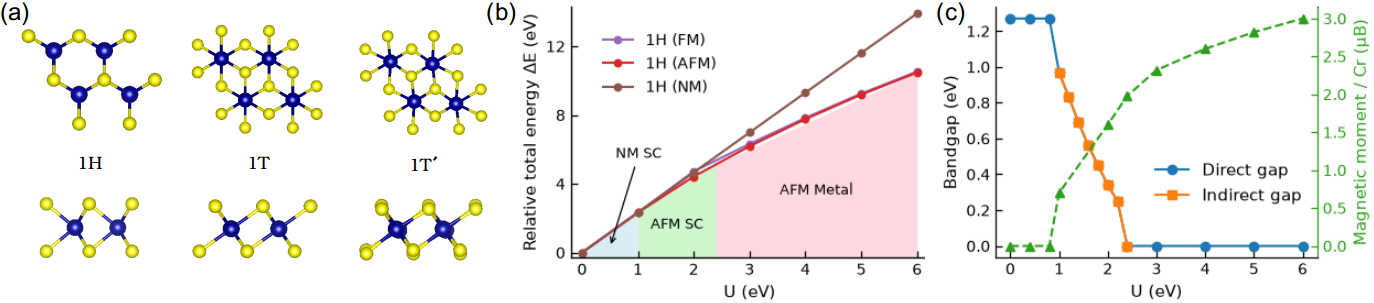}
  \caption{
   \textbf{(a)} Top and side views of relaxed monolayer CrS$_2$ polymorphs (1H, 1T, and 1T$'$), highlighting their distinct coordination geometries. 1T$''$ is omitted, being indistinguishable from 1T at this scale.
   \textbf{(b)} Relative total energy of the 1H phase as a function of the Hubbard U parameter for different magnetic configurations. Shaded regions denote successive electronic and magnetic regimes: non-magnetic semiconductor (NM--SC), antiferromagnetic semiconductor (AFM--SC) and antiferromagnetic metal (AFM--M).
  \textbf{(c)} Evolution of the direct and indirect band gaps (left axis) and local Cr magnetic moment (right axis) with increasing U, illustrating the coupling between gap closure and the emergence of magnetic order.
  }
  \label{Fig. 1}
\end{figure*}

Early first-principles work established 1H-CrS$_2$ as a NM semiconductor with a PBE gap of 1.0--1.3~eV~\cite{Ataca2012,Zhuang2014}.
Zhuang \textit{et al.} further found the 1H structure to lie below 1T with the PBE, HSE06 and van der Waals functionals alike, with no imaginary phonon frequencies and no structural transition in molecular dynamics at 600~K~\cite{Zhuang2014}.
Later studies agree at the semilocal and hybrid levels but diverge as soon as an on-site Coulomb correction is added, and the calculated ground state of CrS$_2$ consequently depends on how that correction is treated.
Chen \textit{et al.} place 1H lowest with both PBE and HSE06: going from PBE to HSE06 roughly doubles the energies of the 1T and 1T$'$ phases relative to 1H, from 0.37 and 0.34 to 0.76 and 0.71~eV/f.u., while leaving the assignments unchanged, with 1H a NM semiconductor of 1.35~eV gap, 1T an antiferromagnetic (AFM) metal, and 1T$'$ a ferromagnetic (FM) semiconductor~\cite{Chen2021}.
On this basis they argued that $U_{\mathrm{eff}}$ should stay below about 1~eV for CrS$_2$, since larger values reverse the ordering~\cite{Chen2021}.
Habib \textit{et al.} find the same pattern with a third code: relative to 1H, the formation energies of the 1T and 1T$'$ phases are $+0.52$ and $+0.37$~eV per CrS$_2$ within GGA and $+0.71$ and $+0.33$~eV within HSE06, but at $U=2.8$~eV the 1T$'$ phase falls to $-0.02$~eV, marginally below 1H~\cite{Habib2019}.
Wang \textit{et al.}, using PBE+U at the same $U=2.8$~eV, reached the opposite conclusion: they found a striped AFM 1T phase more than 20~meV/Cr below the NM 1H phase, an ordering that holds over the whole range $U=1$--5~eV, and concluded that the Hubbard term is needed to recover the magnetic ground state inferred from experiment in Cr chalcogenides~\cite{Wang2018}.
Values of 2.8--3.0~eV are commonly used for this compound~\cite{Wang2018,Kumari2025,Lei2024}.
The disagreement cannot be settled simply by computing $U$ from first principles, since the linear-response value reported for CrS$_2$, 1.5--2.5~eV, already lies in the range where the $+U$ ordering differs from the PBE and HSE06 one~\cite{Wang2018}.

The electronic character of the individual phases is better established.
At the GGA-PBE level including spin--orbit coupling, 1H-CrS$_2$ is a semiconductor with a direct gap of 0.95~eV, 1T is metallic, and 1T$'$ has a fundamental gap of about 32~meV, which those authors describe as semimetallic; the 1H gap is attributed to the filled $d_{z^2}$ level lying below the empty $d_{xy}$ and $d_{x^2-y^2}$ states~\cite{Habib2019}.
The 1T phase is, however, not stable in practice: \textit{ab initio} molecular dynamics at 300~K shows it converting to 1T$'$ by dimerization of the Cr atoms within 6~ps~\cite{Habib2019}, consistent with the unstable phonon branch found for the same structure~\cite{routaray2025}.
\begin{figure*}[!t]
  \centering
  \includegraphics[width=0.95\textwidth]{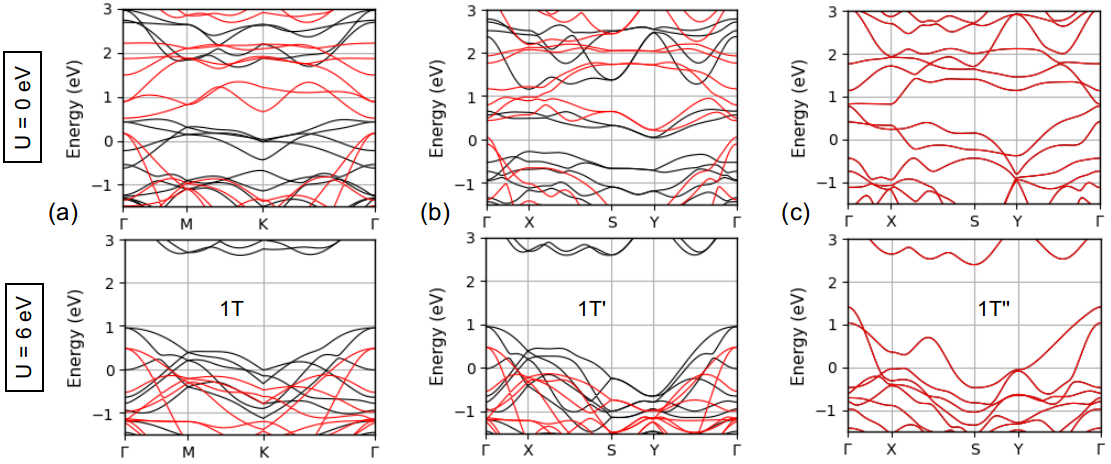}
  \caption{Spin-resolved band structures of the 1T, 1T$'$, and 1T$''$ (a,b,c respectively) phases at representative Hubbard U values (Upper panel U=0eV and bottom panel U=6eV). Increasing U enhances exchange splitting and separates occupied and unoccupied Cr-d states, while the metallic character and magnetic order of these phases remain intact.
  }
  \label{Fig. 2}
\end{figure*}
In 1H-CrS$_2$, compressive strain raises the gap while preserving its direct character~\cite{Zhuang2014}, and the monolayer remains dynamically stable from $-6\%$ to $+6\%$, with a direct-to-indirect transition above 2\% tensile strain~\cite{ChenSB2020}.
Strain has also been used to control magnetic order in the Cr dichalcogenides.
Lv \textit{et al.} showed that biaxial tensile strain switches 1T-CrSe$_2$ from AFM to FM above 2\%, and 1T-CrTe$_2$ from FM to AFM below $-1\%$, and attributed the switch to a competition between the AFM Cr--Cr direct exchange, which weakens quickly as the Cr--Cr distance grows, and the FM $\sim$90$^{\circ}$ Cr--X--Cr superexchange, which does not~\cite{Lv2015}.
For CrS$_2$ itself, Kumari \textit{et al.} applied biaxial strain to the 1T phase at $U=2.8$~eV and reported an AFM-to-FM switch above 2\% strain, a strain-dependent easy axis, and Monte Carlo ordering temperatures of 455~K at zero strain and 516~K at $+8\%$, with a Cr moment of 2.8--3~$\mu_B$ that they ascribe to a mixed Cr$^{3+}$/Cr$^{4+}$ valence produced by S$\rightarrow$Cr charge transfer~\cite{Kumari2025}, whereas the predicted orthorhombic phase retains its FM half-metallic character throughout $-5\%$ to $+5\%$~\cite{Lei2024}.
The response to strain is thus strongly polymorph dependent.
Comparable strain-induced gap reductions and closures have been reported in, e.g., the NM MoS$_2$ and WS$_2$ 2D materials~\cite{Sah2024,Conley2013,keliang2013,Johari2012}.

In this work, we present a first-principles study of the 1H, 1T, 1T$'$, and 1T$''$ structures of monolayer CrS$_2$ under varying Hubbard U and biaxial strain, following the evolution of structural symmetry, magnetic order, and electronic character across all phases. While the 1T-derived structures show largely conventional U-enhanced metallic behavior, the 1H phase passes from a NM semiconductor through an AFM semiconducting regime before becoming metallic.
Resolving the phase diagram as a function of $U$ lets us separate the features of the CrS$_2$ ground state that are stable from those that follow from the choice of $U$, and so compare the assignments of Refs.~\citenum{Chen2021} and~\citenum{Wang2018} directly.
We also extract the magnetic exchange interactions of the 1H phase and follow the Cr--Cr bond disproportionation that accompanies the magnetic transitions.

Two earlier studies are close to the present one and should be distinguished from it.
Chen \textit{et al.} proposed using the polymorph contrast of CrS$_2$ for lateral, strain-controlled spintronics within a single flake, for uniaxially strained 1H, 1T and 1T$'$ phases~\cite{Chen2021}; we apply biaxial rather than uniaxial strain, resolve the dependence on the interaction parameter that they set aside, and extract the exchange interactions, which lets us say where the polymorph contrast comes from rather than reading it off the band structures.
Kumari \textit{et al.} treat the 1T phase at a single value of $U$~\cite{Kumari2025}; here the comparison between polymorphs and the dependence on $U$ are themselves the object of study, and the phase that responds most strongly is 1H.
By clarifying how strain governs phase stability and electronic reconstruction, this work identifies practical routes to strain-driven phase engineering in two dimensions, which is relevant for flexible spintronic devices~\cite{Chen2021FlexibleSpintronics,Serrano2019FlexibleGrapheneSpin,Wang2023SolarPoweredAFMFM}.


\section{Results and Discussions}\label{sec:results}

\subsection{Effect of the Hubbard U parameter on CrS$_2$ properties}\label{subsec:hubbardU}

Figure~\ref{Fig. 1}a shows the relaxed structures of the 1H, 1T and 1T$'$ phases. They differ in the stacking of the two sulfur sublayers across the Cr plane. In the 1H phase ($P\bar{6}m2$) the two sulfur layers lie directly above one another and each Cr atom is coordinated by a trigonal prism of six S atoms, whereas in the 1T phase ($P\bar{3}m1$) one layer is rotated by 60$^{\circ}$ with respect to the other, giving octahedral coordination.

In Figure~\ref{Fig. 1}b we show the evolution of the relative total energy of the 1H phase with increasing Hubbard U for different magnetic configurations. At low U (< 1 eV), the NM configuration of the 1H phase is energetically favored, with only small energy differences between spin configurations, consistent with a NM ground state~\cite{Zhuang2014,Zhang2015DopingCrS2}.
The AFM configuration considered here is the collinear antiparallel arrangement in a 2$\times$1 magnetic supercell.
With increasing U, the AFM configuration becomes energetically favorable, signaling the onset of magnetic ordering driven by enhanced electronic localization. These trends define successive electronic and magnetic regimes, including NM semiconductor (NM--SC), AFM semiconductor (AFM--SC) and AFM metal (AFM--M).
The FM and AFM curves appear nearly overlapping on this scale due to their relatively small energy difference (up to 300 meV) compared to the overall energy variation across U.
A comparison of total energies across different polymorphs is provided in supplementary~\cite{supplm} Figure~S1, where a crossover in phase stability is observed at intermediate U($\approx$ 1--1.5 eV), consistent with earlier studies~\cite{Chen2021,Wang2018}.
Importantly, all phases exhibit smooth and monotonic energy variations with U, indicating the absence of abrupt energetic or structural instabilities and enabling a meaningful comparison of interaction-driven electronic trends.

Figure~\ref{Fig. 1}c shows the evolution of the band gap (direct and indirect) and local Cr magnetic moment with increasing U, clearly delineating distinct U-dependent regimes.
The emergence of a finite magnetic moment precedes the reduction of the band gap, indicating that magnetic ordering develops before the electronic gap begins to close substantially.
At higher U, the gap closes and the system evolves into a metallic phase, highlighting the strong coupling between electronic structure and magnetic order.

\begin{figure*}[!t]
\centering
\includegraphics[width=1\textwidth]{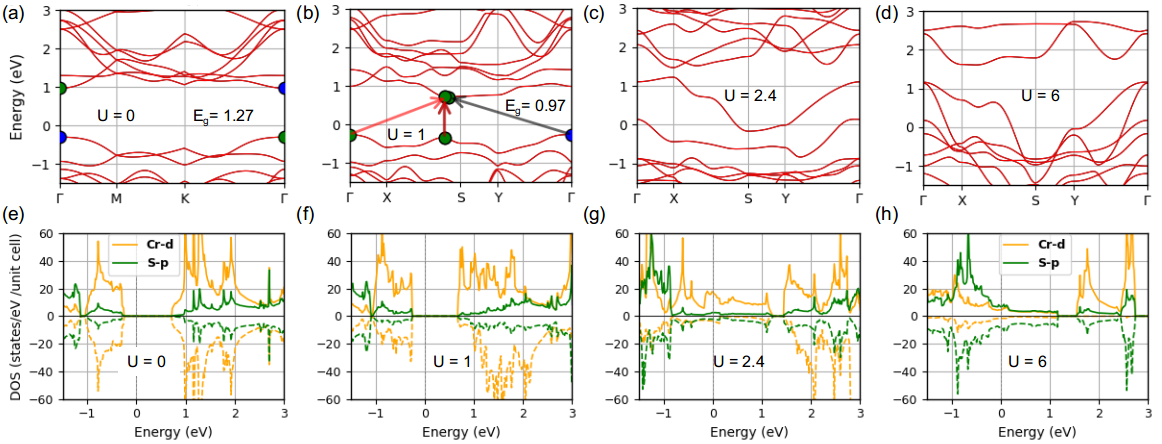}
\caption{\textbf{(a--d)} Band structures of the 1H phase at U = 0, 1, 2.4 and 6 eV, spanning the NM--SC, AFM--SC and AFM--M regimes, showing symmetry lowering, Brillouin-zone folding, band-edge migration, and eventual overlap.
\textbf{(e--h)} Corresponding spin-resolved, $m$-projected density of states per unit cell illustrating the onset of antiferromagnetism and the growth of local Cr magnetic moments. Solid and dashed lines denote spin-up and spin-down channels, respectively.
  }
  \label{Fig. 3}
\end{figure*}

The 1T-derived polymorphs exhibit a largely conventional and robust response to increasing Hubbard U, as illustrated by the band structures in Figure~\ref{Fig. 2} (cf. similar behavior reported for several 2D magnets \cite{Sivadas2015}). For the 1T(FM) and 1T$''$(AFM) phases, increasing U enhances the separation between occupied and unoccupied Cr-d manifolds, leading to the opening of a gap above the Fermi level, while the bands crossing the Fermi energy remain intact. Consequently, both phases retain their metallic character and magnetic order across the full U range, with Hubbard U acting primarily to renormalize higher-energy states rather than inducing a metal--insulator transition.

The 1T$'$(FM) phase displays a slightly richer but still conventional evolution. At low U it is a spin-polarized metal with a low carrier density, one spin channel dominating near the Fermi level. As U increases, the spin asymmetry is reduced and additional band crossings appear, driving a smooth crossover to a fully metallic state, while the underlying FM order remains unchanged.
As DFT+U calculations can exhibit metastable solutions associated with different occupation matrices~\cite{ponet2024}, possible metastability should be kept in mind. Overall, the 1T-derived phases retain their underlying magnetic order, with U primarily modifying the electronic spectrum rather than inducing a qualitative magnetic reconstruction.

By contrast, the 1H polymorph displays a markedly different response to increasing Hubbard U, as shown in the band structure and density of states in Figure~\ref{Fig. 3}(a--h). At U=0, the system is a NM semiconductor with a direct band gap of approximately 1.27 eV at the K-point (Figure~\ref{Fig. 3}(a,e)). The conduction and valence bands are predominantly Cr-$d$ in character, with negligible spin polarization and well-separated spin channels.

Upon increasing U to 1 eV, AFM ordering sets in, evident from the onset of spin asymmetry in the projected density of states (Figure~\ref{Fig. 3}(f)) and the lifting of degeneracies in the band structure (Figure~\ref{Fig. 3}(b)). This transition is accompanied by a reduction in the band gap to approximately 0.97 eV and a notable shift of the conduction and valence band extrema away from high-symmetry points---indicative of symmetry breaking, zone folding, and new $d$-orbital hybridizations. This behavior contrasts with conventional expectations in d-dominated systems, where increased U often widens the gap; here, structural and magnetic symmetry breaking dominates, leading to a gap reduction~\cite{Tan2022,deMedici2011}.

\begin{figure*}[!t]
  \centering
  \includegraphics[width=0.9\textwidth]{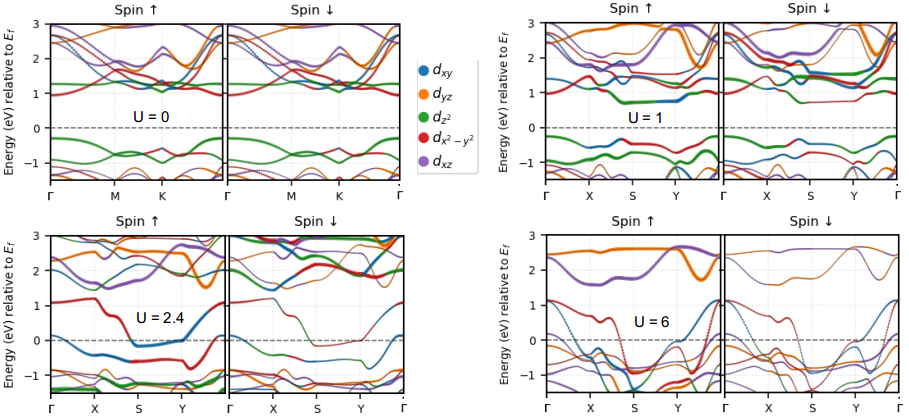}
  \caption{ Orbital-projected fatband representations of the 1H phase at the same critical U values as in Figure~\ref{Fig. 3}, highlighting the gap reduction and closure reflect interaction-induced band reorganization and orbital reshuffling rather than conventional Mott-type gap opening.
  }
  \label{Fig. 4}
\end{figure*}

At U=2.4 eV the gap closes and the system becomes an AFM metal (Figure~\ref{Fig. 3}(c,g)), with valence and conduction bands overlapping at different
k-points. The DOS near the Fermi level becomes increasingly asymmetric, and spin-polarized Cr-$d$ states dominate low-energy excitations. The metallic character becomes more pronounced with further increase of U, and the system remains metallic up to the largest value considered (Figure~\ref{Fig. 3}(d,h)). Importantly, this evolution occurs smoothly, without any abrupt structural instabilities or band inversions, underscoring a continuous interaction-driven transformation rather than a discontinuous phase change~\cite{Imada1998,Zaanen1985}.

At low and intermediate U, the sulfur $p$ states remain well separated from the Fermi level, indicating that the evolution is governed by reorganization within the Cr-$d$ manifold.
The selective enhancement of spin-split $d_{xy}$, $d_{x^2-y^2}$, and $d_{z^2}$ states near the Fermi level signals the emergence of localized magnetic moments and an orbital-dependent spin polarization.
Notably, at higher U, the sulfur $p$-orbitals begin to contribute more significantly near the valence edge---especially in the spin-down channel, indicating growing $p-d$ hybridization and ligand involvement in the low-energy spectrum. This suggests that the metallic regime is shaped not only by Cr-$d$ band reorganization, but also by increased covalency and spin-dependent charge redistribution, consistent with ligand-assisted interaction-driven transitions in low-dimensional materials~\cite{Bennett2021,Grisolia2017}. A further indication of ligand involvement is that the local Cr moment exceeds 3~$\mu_B$ at large U (Figure~\ref{Fig. 1}c), above the spin-only value of 2~$\mu_B$ expected for a $d^2$ Cr$^{4+}$ ion. This points to a partial reduction towards Cr$^{3+}$ accompanied by holes on the sulfur sublattice, as also reported for the 1T phase~\cite{Kumari2025}.
Deciding whether this trend is sufficient to place the metallic state on the charge-transfer side of the Zaanen--Sawatzky--Allen classification~\cite{Zaanen1985} would require the charge-transfer energy $\Delta=\varepsilon_d-\varepsilon_p$ to be evaluated explicitly, which is beyond the scope of the present work. The calculations reported here establish only a progressive growth of the S-$p$ weight at the valence edge with increasing U.

These trends are further clarified in the orbital-resolved fatband plots of Figure~\ref{Fig. 4}, which track the $k$-resolved contributions of individual Cr $d$-orbitals as U increases. At low U, the valence and conduction bands derive primarily from $t_{2g}$-like orbitals with limited orbital mixing and spin asymmetry. However, with increasing U, particularly beyond the onset of magnetic ordering near U $\approx$ 1 eV, a pronounced spin-channel asymmetry emerges: spin-up states gain weight near the Fermi level while spin-down states are progressively suppressed. This evolution signals the formation of localized Cr moments and an increasingly orbital-dependent distribution of spectral weight, most notably in the $d_{x^2-y^2}$ and $d_{z^2}$ orbitals~\cite{deMedici2011}. Concurrently, the symmetry lowering induced by magnetic and structural reordering introduces band folding and hybridization among previously decoupled orbitals. This folding draws valence and conduction band edges closer in momentum space, enabling indirect gap narrowing. This evolution is consistent with the band-gap and magnetic-moment trends shown in Figure~\ref{Fig. 1}c, where gap closure coincides with the emergence of magnetic order.

This highlights that varying U at fixed geometry probes the sensitivity of the electronic structure to orbital localization. In real materials, such localization is influenced by structural factors such as bonding, coordination, and screening, which are inherently environment dependent~\cite{Ghosh2026,Aryasetiawan2004,Ricci2017,Macke2023}.

\subsection{Effect of epitaxial strain}
We now consider biaxial strain as a direct physical route to tune the electronic structure through lattice modification. Biaxial tensile strain provides an experimentally accessible means to control hopping amplitudes, bandwidth, and magnetic exchange pathways by altering interatomic distances and bond angles~\cite{Xu2021,Qi2023,Chaudhuri2023}.
We note that the 1T phase is dynamically unstable as a free-standing monolayer at zero strain and tends to reconstruct to the 1T$'$ phase~\cite{Habib2019,routaray2025}.
We nevertheless consider the strained 1T structure here as a reference configuration to track the effect of lattice deformation on its electronic and magnetic properties.
The strained structures were not subjected to phonon calculations in the present work, so their dynamical stability, particularly beyond the stability range established~\cite{ChenSB2020}, is not assessed here.

\begin{figure}[!t]
\centering
\includegraphics[width=0.95\columnwidth]{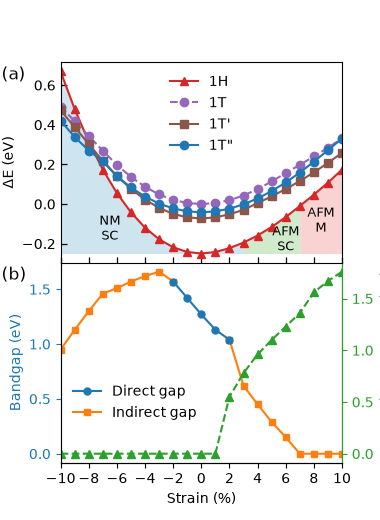}
\caption{
(a) Total energy per formula unit (f.u.) as a function of biaxial tensile strain for all four CrS$_2$ polymorphs. Colored regions on the 1H curve indicate strain-induced electronic and magnetic phases.
(b) Strain dependence of band gap and magnetic moment in the 1H phase, illustrating gap closure and onset of magnetism.
}
\label{Fig. 5}
\end{figure}

\begin{figure*}[!t]
  \centering
  \includegraphics[width=1\textwidth]{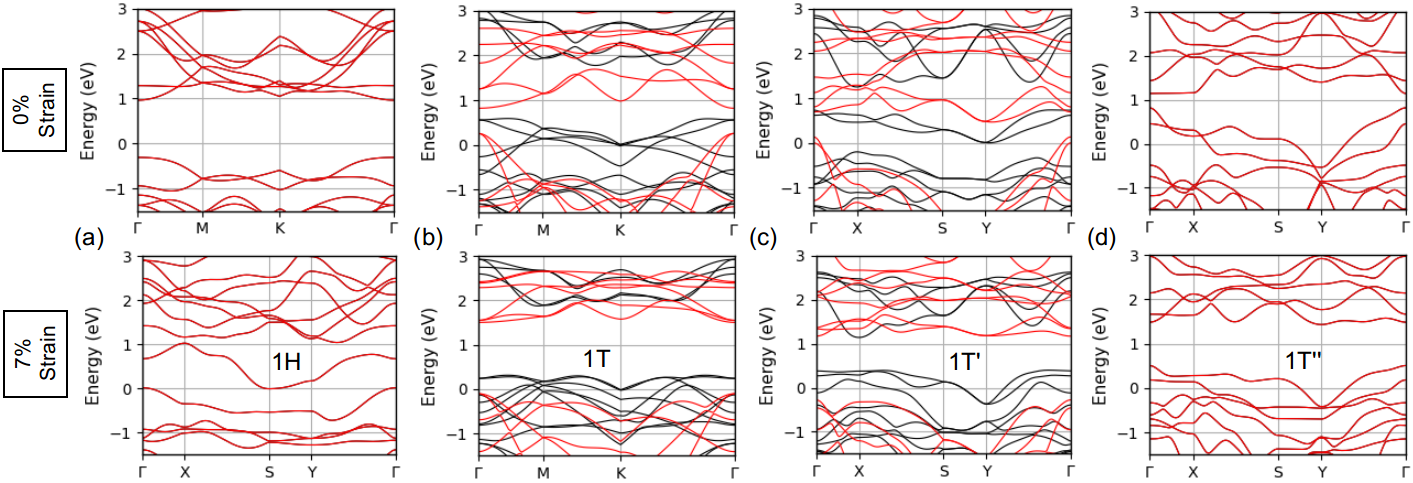}
  \caption{Spin-resolved electronic band structures of the four CrS$_2$ polymorphs (1H(a), 1T(b), 1T$'$(c), 1T$''$(d)) shown in both (upper panel) without and (bottom panel) with biaxial strain(7\%). Red and black colors denote spin up and spin down channels where applicable.}
  \label{Fig. 6}
\end{figure*}

To assess the evolution of phase stability under biaxial strain, we examine the total energy variations of the four CrS$_2$ polymorphs at U=0.6 eV as shown in Figure~\ref{Fig. 5}(a). At zero strain, the 1H phase is the  ground state; however, its energy rises more steeply than the 1T-derived phases under both compressive and tensile strain. Notably, under compressive strain beyond $\sim-8$\%, the energies of the 1T-derived phases fall below that of 1H, suggesting a strain-driven polymorphic crossover. Within the 1H phase, two magnetic transitions are evident: a transition from NM semiconducting to AFM semiconducting behavior near $\sim$2--3\% tensile strain, and a subsequent closure of the gap, giving an AFM metal, at around 7\% similar to the evolution observed upon increasing U in the previous section. For 1H-CrS$_2$, our results reproduce the previously reported strain dependence of the band gap and the direct-to-indirect crossover~\cite{ChenSB2020}, while extending the analysis to larger tensile strains and the accompanying magnetic response.

The impact of strain on the electronic structure is captured more directly in Figure~\ref{Fig. 5}b, where we track the direct and indirect band gaps together with the local Cr magnetic moment of the 1H phase. Under compressive strain, the direct gap initially increases, reaching a maximum before decreasing at larger compression.
With U=0.6 eV, the fundamental gap remains direct between approximately $-2$\% and $+2$\% strain and becomes indirect outside this range. The increasing gap under compressive strain is broadly consistent with the results of Zhuang et al.~\cite{Zhuang2014}. However, our compressive-strain results differ from those of Chen et al.~\cite{ChenSB2020}, who found a direct gap throughout the compressive range, whereas the direct-to-indirect crossover near $+2$\% tensile strain is consistent with their findings.
Under tensile strain, the indirect gap decreases progressively and eventually closes as the system becomes an AFM metal. The finite Cr magnetic moment emerges beyond approximately 2--3\% tensile strain and increases steadily thereafter.

Interestingly, tensile strain produces the same sequence as increasing U. Magnetic order appears first, the
gap narrows, and the system finally becomes metallic.
Similar dual sensitivity of magnetic and electronic behavior to both strain and Hubbard U has also been reported in related Cr-based monolayers~\cite{Zhu2023_CrTe2}.

The microscopic origin of the strain-driven gap closure is apparent in the momentum-resolved band structures of all four polymorphs under 0\% and 7\% tensile strain, as shown in Figure~\ref{Fig. 6}.
Specifically, strain induces band flattening, shifts in the conduction and valence band extrema, and eventually their overlap.
These features arise from strain-driven symmetry breaking and Brillouin-zone reconstruction, which enhance hybridization among Cr-d states and enable gap closure even without invoking additional electronic interaction effects~\cite{Qi2023,Chaudhuri2023}.

By contrast, the 1T-derived polymorphs exhibit no significant qualitative transformation.
Their metallic character remains robust under applied strain, which predominantly alters bandwidths and band dispersions without triggering gap opening or magnetic reconstruction.
This stability highlights the uniquely enhanced susceptibility of the 1H phase to strain-driven electronic reorganization under external perturbations.
The orbital-resolved bands in Supplementary~\cite{supplm} Figure~S2 show a similar reorganization of the Cr-d-derived states near the Fermi level to that observed with increasing U in Figure~\ref{Fig. 4}.
In both cases, the band extrema shift in momentum and the $d_{xy}$, $d_{z^2}$, and $d_{x^2-y^2}$ states acquire increased weight near the Fermi level as the system approaches the metallic regime.
Together with Figure~\ref{Fig. 5}, these results show that biaxial strain can drive symmetry breaking, electronic phase transitions, and magnetic reconstruction, particularly under tensile strain in the 1H polymorph.

In the 1H phase, strain induces a sequence of magnetic and electronic transitions by tuning the bandwidth and exchange pathways, thereby providing a direct route to tune the balance between localized and itinerant electronic character.

\subsection{Exchange Interactions and Magnetoelastic Coupling}
To establish a microscopic connection between the interaction- and strain-driven phase transitions and the underlying magnetic interactions, we extract effective exchange parameters for the 1H phase using the following Heisenberg spin Hamiltonian,

\begin{equation}
\mathcal{H}
=
-\sum_{i \neq j}
J_{ij}\,
\mathbf{S}_i \cdot \mathbf{S}_j ,
\qquad
J_{ij} \in \{J_1, J_1', J_2, J_2', J_3, J_3'\},
\label{eq:Heisenberg}
\end{equation}

where $J_{ij}$ denotes the exchange interaction between Cr moments at sites $i$ and $j$.

As a consequence of reduced symmetry induced by electronic interactions (via U) and strain, crystallographically equivalent Cr--Cr pairs in the high-symmetry phase split into inequivalent exchange channels, denoted as $J_1, J_1', J_2, J_2'$ and $J_3, J_3'$ corresponding to distinct nearest- and next-nearest-neighbor pathways \cite{Lado2017,Sivadas2015}.

\begin{figure*}[!t]
  \centering
  \includegraphics[width=1\textwidth]{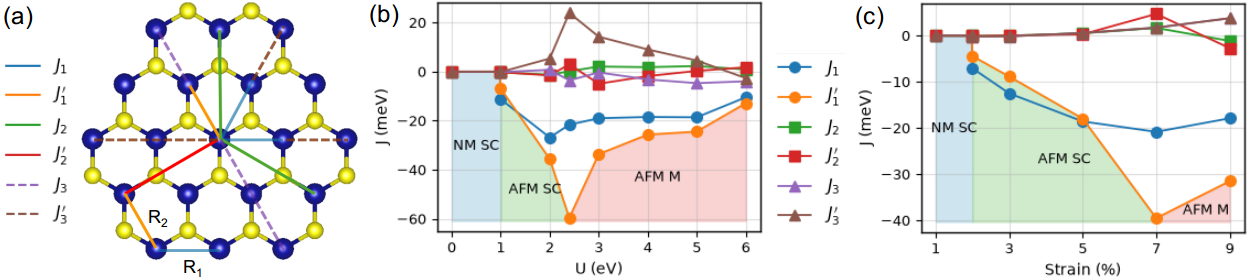}
  \caption{\textbf{(a)} Cr--Cr exchange paths and bond lengths used in Heisenberg mapping.
  \textbf{(b,c)} Exchange parameters $J_i$ as functions of Hubbard U and biaxial strain(at U=0.6 eV) for 1H phase, respectively, highlighting the emergence of bond-dependent AFM exchange and the crossing of $J_1$ and $J'_1$. Colored regions denote the NM--SC, AFM--SC, and AFM--M regimes.
  }
  \label{Fig. 7}
\end{figure*}

The schematic in Figure~\ref{Fig. 7}(a) shows the real-space Cr--Cr magnetic interaction network: nearest-neighbor ($J_1$, $J_1'$), second-neighbor ($J_2$, $J_2'$), and third-neighbor ($J_3$, $J_3'$) paths, along with the corresponding bond lengths $R_1$ and $R_2$.
Electronic interactions (as modeled by U) and strain both lead to symmetry lowering, making these exchange channels inequivalent and magnetically anisotropic~\cite{Burch2018, Lado2017}.

Figure~\ref{Fig. 7}(b) presents the evolution of the exchange interactions with increasing Hubbard U. At low U, the 1H phase is NM and has no well-defined local moments and, consequently, the Heisenberg mapping is not applicable in this regime, and we set the exchange parameters $J_{ij}$=0 by construction.
In the \textsc{SIESTA} data reported in the Supplementary Material~\cite{supplm} the corresponding couplings are instead computed explicitly and are found to vanish wherever the local moment vanishes, confirming this choice.  Once finite magnetic moments emerge above U $\approx$ 1 eV, the exchange parameters become well defined and evolve with U.
As U exceeds the threshold for AFM order ($\sim$1 eV), $J_1$ and $J_1'$ diverge significantly, with $J_1'$ becoming strongly AFM.
Higher-order couplings ($J_2$, $J_3$ and their primed counterparts) also evolve, becoming inequivalent and exhibiting non-monotonic behavior due to orbital rehybridization and emerging symmetry lowering.
Notably, $J_1$ and $J_1'$ cross at intermediate U, indicating a redistribution of the dominant exchange pathways.

This reorganization correlates with the sequence of electronic phases from a NM state to an AFM semiconductor and eventually an AFM metal. We note already at this point that a localized Heisenberg description becomes questionable once the gap closes, so the couplings plotted beyond $U\approx2.4$~eV should be taken as indicative of the trend rather than as quantitative exchange constants; this is examined further in the comparison with \textsc{SIESTA} at the end of this section.

\begin{figure}[!t]
  \centering
  \includegraphics[width=0.99\columnwidth]{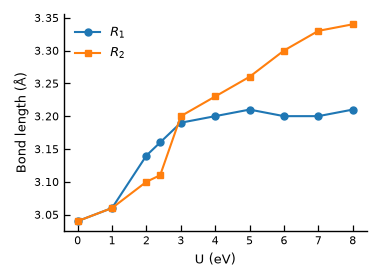}
  \caption{Evolution of inequivalent Cr--Cr bond lengths $R_1$ and $R_2$ with U, demonstrating interaction-driven bond disproportionation (breathing-like distortion).
  }
  \label{Fig. 8}
\end{figure}

The evolution of exchange interactions is primarily driven by interaction-induced electronic reorganization.
As U increases, changes in orbital occupation and hybridization modify the competition between exchange channels, leading to the observed crossover in $J$ even in the absence of structural relaxation.
Structural relaxation introduces an additional, cooperative effect through a U-driven bond disproportionation, as illustrated in Figure~\ref{Fig. 8}.
The initially equivalent Cr--Cr bonds split into shorter ($R_1$) and longer ($R_2$) subgroups, with increasing U amplifying this disparity.
This breathing-type distortion, analogous to that observed in rare-earth nickelates~\cite{Johnston2014,Hepting2020,Bisogni2016}, enhances the inequivalence of hopping amplitudes and further differentiates the exchange pathways.

The development of local magnetic moments with increasing U is shown in Figure~\ref{Fig. 1}c, where the onset and growth of magnetization follow the same U-dependent thresholds governing the gap closure and exchange reorganization.

We now turn to strain as an independent physical control parameter.
Figure~\ref{Fig. 7}(c) shows the exchange parameters as functions of biaxial tensile strain (at fixed $U=0.6$ eV).
Similarly to the U-driven case, strain induces a growing disparity between $J_1$ and $J_1'$, leading to a crossover in their strength.
Higher-order interactions ($J_2$, $J_3$ and their primed analogues) also undergo strain-driven evolution, indicating that lattice deformation alone can tune the dominant magnetic pathways.
The strain dependence of the magnetic moment in Figure~\ref{Fig. 5}(b) mirrors the trend observed with increasing Hubbard U, with AFM order emerging beyond $\approx$ 2\% strain and the moment increasing steadily thereafter.

Together, these results highlight strong magnetoelastic coupling in 1H--CrS$_2$: both electronic interactions and lattice strain promote bond distortions and magnetic order through distinct yet converging routes.
Magnetic order already develops when the atomic positions and the cell are held at the $U=0$ relaxed geometry, under both perturbations, indicating that the transition originates from an electronic instability rather than from structural relaxation.
While Hubbard U directly enhances on-site electronic interactions and promotes localization, strain perturbs the electronic structure through lattice deformation, effectively shifting the balance between competing exchange pathways.
Although the underlying microscopic mechanisms differ, both strain and Hubbard U correction lead to a comparable reorganization of exchange interactions and magnetic symmetry breaking.
Structural relaxation further stabilizes the ordered phase but plays a secondary role in its onset.
The strain- and U-tunable exchange landscape thus positions CrS$_2$ as a versatile platform for magnetic phase engineering, with the similar evolution of exchange interactions under these distinct perturbations.

To assess the robustness of the exchange interactions underlying the magnetoelastic picture discussed above, we performed independent calculations using \textsc{TB2J} directly on the localized-orbital Hamiltonian obtained from \textsc{SIESTA}, without intermediate Wannierization.
This provides an independent comparison with the \textsc{ABINIT} results, particularly near the electronic gap closure, where the Wannier representation becomes more difficult to control. The full comparison is presented in Supplementary~\cite{supplm} Section~S3.

In the $2\times1$ magnetic cell, $J_1$ couples antiparallel spins, whereas $J_1'$ connects parallel spins along $\mathbf{a}_2$.
Although these bonds are equivalent in the undistorted NM lattice, magnetic ordering and structural relaxation differentiate their local environments and Cr--Cr distances.

The \textsc{SIESTA} calculations reproduce the main qualitative trends, with both couplings becoming AFM upon the formation of local moments and $J_1'$ increasing more rapidly than $J_1$, eventually becoming dominant. The crossing occurs
between $U=1.0$ and $1.2$~eV, consistent with the \textsc{ABINIT} results presented above, and between $+2$ and $+3\%$ strain, i.e. earlier than found with the \textsc{ABINIT} calculations.
The associated bond disproportionation is also reproduced, with the Cr--Cr distances approaching approximately $3.20$ and $3.34$~\AA.
The earlier onset under strain is consistent with a general displacement of the magnetic and electronic thresholds towards lower $U$ and lower strain in the localized-orbital calculations, which may reflect differences in the correlated-subspace representation and other aspects of the two implementations, and indicates that the transition thresholds are method dependent.

The comparison also reveals that the nearest-neighbour couplings do not increase monotonically throughout the entire $U$ range, and that the two channels do not turn over together.
In \textsc{SIESTA}, $J_1$ extremises near $U\approx1.2$~eV and $J_1'$ near $U\approx1.5$~eV, reaching $-28$ and $-49$~meV respectively (negative $J$ denotes AFM coupling in the convention of Eq.~\ref{eq:Heisenberg}, as plotted in Figure~\ref{Fig. 7}), to be compared with $-36$ and $-60$~meV at $U\approx2.0$ and $2.4$~eV in the \textsc{ABINIT} calculations, and both weaken thereafter even though the local Cr moment continues to grow monotonically.
The separation of the two extrema is structural rather than magnetic in origin: above $U\approx1.5$~eV the parallel-spin bond expands by more than $0.25$~\AA\ while the antiparallel bond remains essentially unchanged, so the ligand-mediated coupling on the former is progressively cut off, whereas the latter weakens only gradually with an additional drop at the gap closure.
The same ordering is found under strain, the two channels reaching their largest magnitudes
near $+5$ and $+6\%$.
Beyond the gap closure the two methods no longer agree on the sign of the parallel-spin coupling, which highlights the limitations of a localized Heisenberg mapping in the metallic regime, and we therefore restrict our quantitative interpretation to the gapped regime.
This limitation applies equally to the \textsc{ABINIT} couplings discussed earlier in this section: the values shown beyond the gap closure in Figure~\ref{Fig. 7}(b) are reported for completeness and indicate the trend, but should not be read as quantitative exchange constants.
The \textsc{SIESTA} calculations further show that compressive strain suppresses the local moment entirely below about $-3\%$, reducing all exchange interactions to negligible values.

Finally, including the sulfur contribution modifies $J_1$ by only 1--6\% in the gapped regime, compared with about 10\% for $J_1'$, enhancing its AFM character. This difference points to a stronger ligand-mediated contribution to $J_1'$, consistent with the increasing sulfur-$p$ character near the valence-band edge discussed in Section~\ref{subsec:hubbardU}.
The complete exchange parameters and ligand-correction analysis are provided in Supplementary~\cite{supplm} Figures.~S5 and~S7.


\section{Conclusion}
\label{sec:conclusion}
We have shown that monolayer CrS$_2$ is a tunable two-dimensional material in which electronic interactions, magnetism, and lattice degrees of freedom are strongly coupled.
By systematically varying the on-site Hubbard interaction and biaxial tensile strain across multiple structural polymorphs, we uncover a pronounced phase dependence of interaction-driven effects~\cite{Huang2021,Jang2023}.

The 1T-derived phases (1T, 1T$'$, and 1T$''$) display largely conventional behavior, remaining metallic with enhanced exchange splitting as the Hubbard U increases~\cite{Li2021tmdreview}.
In contrast, the 1H phase undergoes a sequence of transitions from a non-magnetic semiconductor to an antiferromagnetic semiconductor and finally a metal.
This evolution is driven by interaction-induced orbital and band reconstruction of Cr-d--derived states, accompanied by symmetry lowering and Brillouin-zone reorganization, ultimately leading to progressive gap reduction and closure rather than the conventional localization-driven gap opening expected in simple Mott scenarios~\cite{Wang2023metaltoinsulator}.
The observed metallization in the 1H phase does not follow a simple bandwidth-controlled Mott scenario,
but is instead accompanied by an orbital-dependent redistribution of spectral weight among the Cr-$d$ states. Whether this is related to the orbital-selective behaviour discussed for multiorbital correlated systems~\cite{Yu2017OSMP,Song2021OSnFL} cannot be decided within a static mean-field treatment such as DFT+U, which gives no access to orbital-resolved quasiparticle weights.

We further identify strong magnetoelastic coupling in the 1H phase, manifested as a static breathing-type Cr--Cr bond disproportionation that reorganizes magnetic exchange interactions and accompanies the observed magnetic and electronic transitions.
Importantly, biaxial strain reproduces similar phase evolution by modifying the electronic structure and exchange pathways, demonstrating that lattice perturbations can access the same underlying instability through a distinct physical route~\cite{Zhao2023straincontrol}.

The sequence of magnetic and electronic transitions reported here, and the accompanying reorganization of the exchange interactions, were reproduced with an independent electronic-structure method (SIESTA), which indicates that the phase diagram is transferable between different implementations and approximations even though the critical values of $U$ and of strain at which the transitions occur are not (Supplementary~\cite{supplm} Section~S3).

While synthesizing monolayer CrS$_2$, achieving a single structural phase can be experimentally challenging due to phase mixing, defects, or growth-induced perturbations---as evidenced by the coexistence of multiple phases of ML CrS$_2$ in experiments~\cite{Habib2019}.
Nevertheless, our results indicate that the 1H phase plays a dominant role in U- and strain-induced transitions, whereas 1T-derived polymorphs remain comparatively insensitive.
This suggests that functional switching behavior can still be realized in mixed-phase samples, provided the 1H phase is present.

The polymorph contrast reported here also suggests a perspective that we have not pursued quantitatively. Because the 1H phase changes its magnetic and electronic character under strain while the 1T-derived phases do not, spatial variations of strain or of the local environment could in principle stabilize coexisting domains of distinct character within a single monolayer, giving atomically coherent boundaries between semiconducting and metallic regions without the disorder introduced by chemical doping~\cite{Gong2019,Cheng2022,Liu2022_lateral}. Single-material lateral heterostructures of this kind have already been proposed for CrS$_2$ on the basis of uniaxial strain~\cite{Chen2021}, and the biaxial phase diagram obtained here indicates that the same route should remain open under biaxial loading. Testing it would require explicit interface calculations, including the energetics and width of the domain wall itself, which lie beyond the scope of the present work.
These features make CrS$_2$ and related chromium dichalcogenides promising candidates for flexible spintronic devices that exploit sharp phase boundaries and tunable magnetic exchange~\cite{Manzeli2017,Zhao2021}.

Moreover, our results provide microscopic insight into how electronic interactions and lattice effects cooperate, laying the groundwork for future studies exploring doping, gating, and other external stimuli in correlated Cr-based two-dimensional systems.
The present findings may also have broader implications beyond CrS$_2$, as the C2DB~\cite{Haastrup2018,Gjerding2021} database reports at least 82 two-dimensional materials with non-zero band gap at U = 0 eV and zero band gap at U = 4 eV, along with several others exhibiting band-gap reduction upon increasing the Hubbard U.
The microscopic picture developed here may therefore help rationalize the origin of such unconventional behavior in this broader family of correlated materials.


\section{Computational details}
The first-principles calculations presented in this work were carried out using the \textsc{ABINIT} software package~\cite{Gonze2020,Romero2020} (version 10.3.6).
The interaction between valence electrons and ionic cores was described using the projector augmented-wave (PAW) method~\cite{Blochl1994PAW, Torrent2008}, with pseudopotentials taken from the PseudoDojo project~\cite{VanSetten2018} (version~0.5) and the generalized gradient approximation in the Perdew--Burke--Ernzerhof (GGA-PBE) form for the exchange-correlation functional~\cite{Perdew1996}.
Spin--orbit coupling is neglected in the present calculations.
For monolayer CrS$_2$, previous SOC-inclusive calculations report valence-band splittings of 69.8 meV at the K point~\cite{Habib2019} and approximately 74 meV~\cite{Wang2018}.
These energy scales are small compared with the electronic reconstructions examined here, which occur over several hundred meV to eV, and SOC is therefore omitted in the present analysis.
To ensure consistency across all calculations, a uniform scheme for Brillouin zone sampling was employed for all systems.
The plane-wave kinetic energy cutoff was set to 36~Ha (approximately 980~eV), with a corresponding PAW augmentation cutoff \texttt{pawecutdg} of 72~Ha.
Brillouin zone integrations were performed using an $10\times20\times1$ Monkhorst--Pack \emph{k}-point mesh for all materials studied.
Structural relaxations were performed until the residual Hellmann--Feynman forces on each atom were smaller than $10^{-8}$~Ha/Bohr (approximately $5.1\times10^{-7}$~eV/\AA).
A vacuum spacing exceeding 15~\AA\ was introduced along the out-of-plane direction to eliminate spurious interactions between periodic images.
Electronic occupations were treated using a Gaussian smearing width of 0.002~Ha (about 0.054~eV)

Electronic interactions were treated within the GGA+U framework~\cite{Anisimov1991,Anisimov1997}.
To systematically assess the sensitivity of the structural, magnetic, and electronic properties to the choice of U, we performed calculations over a broad range of U=0--8 eV.

We find that for U<1 eV, the computed ground-state properties are consistent with the previously reported stability of the 1H phase and its non-magnetic semiconducting character~\cite{Chen2021,Habib2019}.
The 1T$'$ phase remains FM and is a
spin-polarized metal with a low carrier density, closer to the semimetallic character reported in Ref.~\citenum{Habib2019} than to the FM semiconducting state of Ref.~\citenum{Chen2021}. Deviations from the U=0 electronic structure become noticeable only above U $\approx$ 0.8 eV.
Because the calculated properties show little qualitative variation over U = 0--0.8 eV, we adopt U = 0.6 eV as a representative value within this low-U regime for the strain-dependent calculations.
Using a common U across all polymorphs allows a consistent comparison of their strain response while preserving the non-magnetic ground state of 1H-CrS$_2$ at zero strain.

The magnetic exchange parameters were computed using the \textsc{TB2J}~\cite{He2021TB2J} package, which extracts Heisenberg exchange interactions from first-principles calculations based on Wannier functions. Wannier functions were constructed separately using the same orbital (Cr-d and S-p) projections, while the disentanglement and frozen energy windows were adjusted for each U and strain values. Across all calculations, the outer disentanglement-window boundaries varied from $-7.2$ to $-6.7$ eV at the lower edge and from 3.2 to 4.3 eV at the upper edge; the corresponding frozen-window boundaries varied from $-7.0$ to $-6.65$ eV and from 3.0 to 4.25 eV, respectively, relative to the Fermi level.

Since the Hubbard correction is not applied to a uniquely defined set of orbitals, and since several of the transitions discussed above involve energy differences of only a few tens of meV per formula unit, the robustness of the results with respect to the electronic-structure
method was verified by repeating the calculations with a second, independent code.
These complementary calculations were performed with \textsc{SIESTA}~\cite{soler2002siesta,gomez2023compatibility},
which employs numerical atomic
orbitals rather than plane waves, and norm-conserving pseudopotentials rather than the PAW method~\cite{Blochl1994PAW,Torrent2008}, so that the Hubbard term acts on a localized atomic orbital rather than on the PAW partial waves.
Since the on-site exchange parameter of the Hubbard correction is set to zero in both cases, the two implementations reduce to the same simplified rotationally invariant
functional, so that the projector is the principal remaining difference in the treatment of $U$.
The same PBE functional, the same $2\times1$ magnetic cells and the same values of $U$ were used throughout, with $U$ applied to the Cr-3$d$ shell only.
Fully relativistic, norm-conserving pseudopotentials were taken from the Pseudo-Dojo database~\cite{VanSetten2018} in PSML format~\cite{garcia2018psml}.
A double-$\zeta$ polarized (DZP) basis was used, together with a mesh cutoff of 1000~Ry and a $12\times24\times1$ Monkhorst--Pack $k$-point mesh.
All structures were relaxed again within this scheme.
Magnetic exchange parameters were obtained with the LKAG method as implemented in \textsc{TB2J}~\cite{He2021TB2J}, applied directly to the \textsc{SIESTA} Hamiltonian and therefore requiring no intermediate Wannier construction, including both Cr and S and subsequently downfolding the ligand contribution onto the Cr--Cr couplings.
The full comparison between the two methods is reported in Supplementary~\cite{supplm} Section~S3.

\begin{acknowledgments}
The authors thank Xavier Gonze and He Xu for their insightful discussions.
This research was carried out within the framework of the Shapeable 2D Magnetoelectronics by Design (SHAPEme, EOS Project No. 40007525) initiative, which is supported by the Fonds de la Recherche Scientifique (FWO) and the FRS-FNRS through the Belgian Excellence of Science (EOS) program. Computational resources were provided by the high-performance computing facilities of the Consortium des \'{E}quipements de Calcul Intensif (CECI), funded by the FRS-FNRS under Grant No. 2.5020.11.
\end{acknowledgments}

\bibliographystyle{apsrev4-2}
\bibliography{rsc}

\end{document}